\documentclass[sn-nature]{sn-jnl}

\usepackage{aas_macros}
\usepackage{siunitx}

\usepackage{graphicx}%
\usepackage{multirow}%
\usepackage{amsmath,amssymb,amsfonts}%
\usepackage{amsthm}%
\usepackage{mathrsfs}%
\usepackage[title]{appendix}%
\usepackage{xcolor}%
\usepackage{textcomp}%
\usepackage{manyfoot}%
\usepackage{booktabs}%
\usepackage{algorithm}%
\usepackage{algorithmicx}%
\usepackage{algpseudocode}%
\usepackage{listings}%

\title{From proto-neutron star dynamo to low-field magnetars}

\author*[1]{\fnm{Andrei} \sur{Igoshev}}\email{a.igoshev@leeds.ac.uk} 
\author[2]{\fnm{Paul} \sur{Barr\`{e}re}} 
\author[3]{\fnm{Rapha\"{e}l} \sur{Raynaud}}
\author[2]{\fnm{J\'{e}rome} \sur{Guilet}} 
\author[4]{\fnm{Toby} \sur{Wood}}
\author[1]{\fnm{Rainer} \sur{Hollerbach}}

\affil*[1]{\orgdiv{Department of Applied Mathematics}, \orgname{University of Leeds}, \city{Leeds}  \postcode{LS2 9JT}  \country{UK}}
\affil[2]{\orgname{Universit\'{e} Paris-Saclay}, Universit\'{e} Paris Cit\'{e}, CEA, CNRS, AIM,  \postcode{91191} \city{Gif-sur-Yvette},  \country{France}}
\affil[3]{\orgname{Universit\'{e} Paris Cit\'{e}}, Universit\'{e} Paris-Saclay, CEA, CNRS, AIM,  \postcode{F-91191}, \city{Gif-sur-Yvette},  \country{France}}
\affil[4]{\orgdiv{School of Mathematics, Statistics and Physics}, \orgname{Newcastle University}, \city{Newcastle upon Tyne}  \postcode{NE1 7RU}  \country{UK}}

\date{}

\begin{document} 





\abstract{
Low-field magnetars have dipolar magnetic fields that are 10-100 times weaker than the threshold, $B \gtrsim 10^{14}$~G, used to define classical magnetars, yet they produce similar X-ray bursts and outbursts. Using the first direct numerical simulations of magneto-thermal evolution starting from a dynamo-generated magnetic field, we show that the low-field magnetars can be produced as a result of a Tayler--Spruit dynamo inside the proto-neutron star. We find that these simulations naturally explain key characteristics of low-field magnetars: (1) weak ($\lesssim 10^{13}$~G) dipolar magnetic fields, (2) strong small-scale fields, and (3) magnetically induced crustal failures producing X-ray bursts. These findings suggest two distinct formation channels for classical and low-field magnetars, potentially linked to different dynamo mechanisms.
}


\maketitle

Magnetars play a special role in modern high-energy astrophysics. They were suggested as central engines for superluminous supernovae \cite{Wheeler2000ApJ,Dessart2012MNRAS} and ultra-long $\gamma$--ray bursts \cite{Greiner2015Natur}. They 
produce at least a fraction of mysterious Fast Radio Bursts \cite{PopovPostnov2010vaoa,Bochenek2020Natur}. While Galactic magnetars are scarce due to their short life --- with 30 known magnetars, compared with 3500 radio pulsars based on ATNF catalogue v.2.1.1\footnote{http://www.atnf.csiro.au/research/pulsar/psrcat} \cite{ATNF2005AJ} --- it is estimated that around $10$\% of all neutron stars (NSs) 
undergo a magnetar stage at some point in their evolution \cite{KeaneKramer2008MNRAS}.

The standard magnetar model explains quiescent X-ray emission, spin period, bursts, outbursts and giant flares observed from Anomalous X-ray Pulsars (AXP) and Soft Gamma Repeaters (SGR) by assuming that these NSs have strong dipolar magnetic fields $\gtrsim 10^{14}$~G \cite{ThompsonDuncan1993ApJ,Thompson1995MNRAS}.
However, a significant fraction of magnetars (5 out of 30 known objects) in fact have dipolar magnetic fields well below $10^{14}$~G and were therefore named low-field magnetars \cite{vanderHorst2010ApJ, Rea2010Sci,Rea2012ApJ,Scholz2012ApJ,Rea2014ApJ}. It has been suggested that low-field magnetars are old neutron stars primarily powered by crust-confined toroidal magnetic fields with strength $\approx 10^{14}$~G \cite{Rea2010Sci,Igoshev2021NatAs}. Rea et al.~\cite{Rea2012ApJ} suggested that low-field magnetars were born with both poloidal and toroidal magnetic fields $>10^{14}$~G, but the poloidal component decayed by a factor of six in $\approx 500$~kyr. Phase-resolved X-ray observations show that in two cases low-field magnetars host small-scale magnetic fields which are 10-100 times stronger than their dipolar fields \cite{Tiengo2013Natur, RodriguezCastillo2016MNRAS}.

The origin of magnetar magnetic fields is a subject of debate \cite{Sarin2023ApJ}. Different dynamo mechanisms have been proposed to explain the formation of the strongest magnetic fields, including proto-neutron star convection \cite{ThompsonDuncan1993ApJ,Raynaud2020SciA,Raynaud2022MNRAS,Masada2022,White2022}, magnetorotational instability \cite{ReboulSalze2021AA,ReboulSalze2022AA}, and more recently the Tayler--Spruit dynamo \cite{Spruit2002AA,Barrere2022AA,Barrere2023MNRAS}. The Tayler--Spruit dynamo is a particularly promising mechanism for generating magnetars' magnetic fields in cases when the progenitor core is slowly rotating and the proto-NS is spun up by fallback accretion \cite{Barrere2022AA}.
In cases of rotation periods slower than ten milliseconds, a normal core-collapse supernova is expected to occur, in agreement with observational constraints for the majority of magnetars \cite{Vink2006,Martin2014}. After the first minute, the proto-NS cools down, its crust solidifies and the remnant becomes a NS. After this time, the initially complicated crustal magnetic field slowly relaxes due to Ohmic decay and Hall evolution on a timescale of $10^5$--$10^6$ years \cite{GoldreichReisenegger1992ApJ,Igoshev2021Univ}.

Previous simulations of magneto-thermal evolution have assumed idealised initial conditions rather than magnetic configurations generated by a specific dynamo mechanism. However, the study of more realistic initial conditions is of key importance in order to obtain realistic predictions of magnetar properties. Indeed, Hall evolution has been shown to preserve certain aspects of the initial conditions \cite{Hollerbach2004MNRAS,WareingHollerbach2010JPlPh}. Hence, the observational properties of magnetars, and low-field magnetars in particular, should contain information about the proto-NS magnetic field.

\section*{Evolution of neutron star magnetic field}

The proto-NS dynamo and NS crust stages are modelled separately because of their very different timescales and physical conditions. While the dynamo is formulated as a magnetohydrodynamics (MHD) problem for a stably stratified fluid with shear caused by fallback accretion over a timescale of a few tens of seconds, the magneto-thermal evolution of the NS crust occurs on a much longer timescale of 1~Myr and is formulated as electron-MHD.   

\begin{figure*}
\includegraphics[width=0.99\columnwidth]{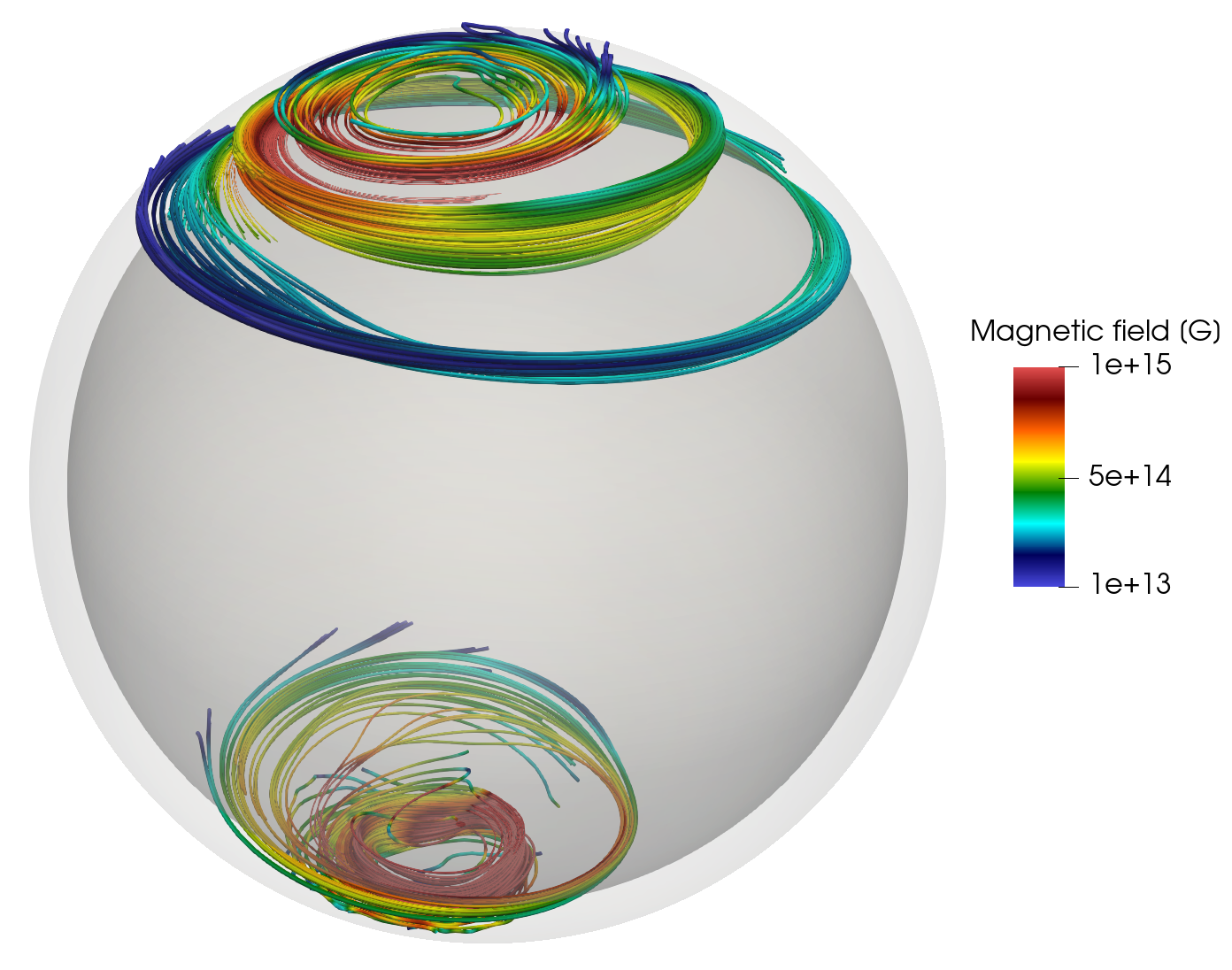}
    \caption{
    Magnetic field lines at the beginning of our NS magneto-thermal simulation.
    }
    \label{fig:A1A2_init}
\end{figure*}

The initial condition for our NS simulation is a magnetic field configuration corresponding to a Tayler-Spruit dynamo branch recently discovered in direct numerical simulations and characterised by a dipolar symmetry (i.e. equatorially symmetric) \cite{Barrere2023MNRAS}. The initial core temperature is assumed to be $10^8$~K.
This magnetic field is obtained using the 3D spherical MHD code \texttt{MagIC}~\cite{Wicht2002PEPI,Gastine2012Icarus,Schaeffer2013GGG} for rotation frequencies of the respective outer and inner spheres $\Omega_o=4\Omega_i=628$~rad$\,$s$^{-1}$ (see Methods Section \ref{s:protons} for a more detailed description).
The magnetic field is predominantly toroidal and reaches values up to $3\times 10^{15}$~G inside the volume, but the field at the outer boundary is much weaker. Assuming a scenario in which the core magnetic field is expelled to a crust-confined configuration, we extract the magnetic field in the top 10\% of the simulation volume and adapt it to our code to model crust-confined NS magneto-thermal evolution (see Methods Section~\ref{a:conversion}, \ref{sm:configuration}). Figure~\ref{fig:A1A2_init} shows
the initial configuration of the magnetic field inside the NS crust.
We then use the \texttt{PARODY} code \cite{dormy1997,parody1,parody2} to integrate the coupled magnetic induction and thermal diffusion equations for 1~Myr before analyzing the NS magnetic characteristics (see Methods Section~\ref{sm:magnetothermalsimulations}).

\begin{figure}
	\includegraphics[width=1\columnwidth]{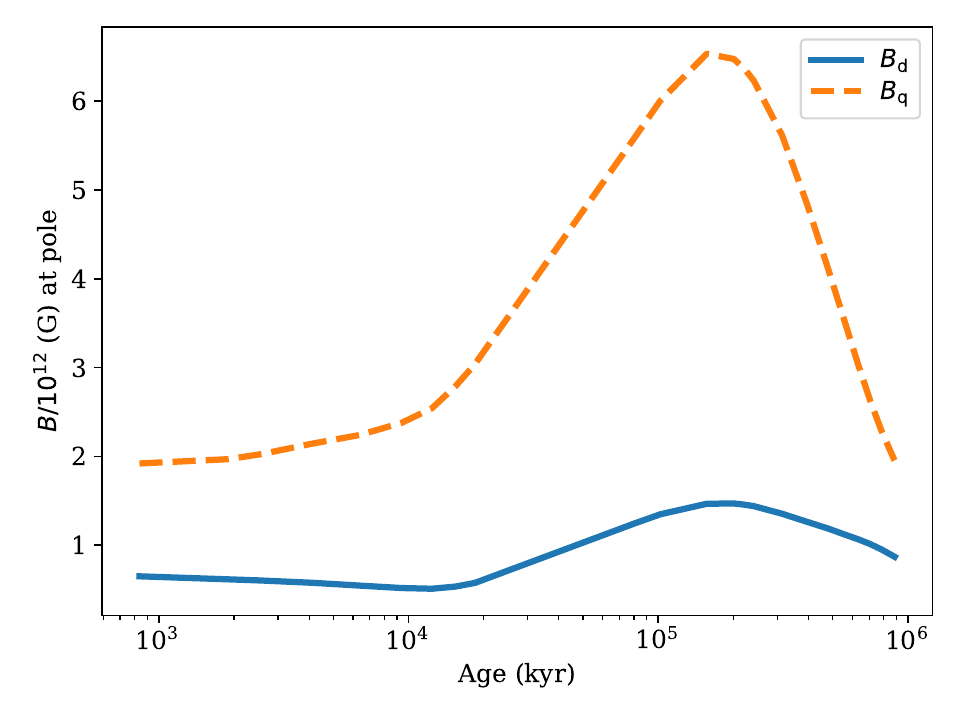}
    \caption{Evolution of surface dipole (blue solid line) and quadrupole (orange dashed line) magnetic fields. 
    }
    \label{fig:magnetconf}
\end{figure}

Figure~\ref{fig:magnetconf} shows the dipolar and quadrupolar poloidal magnetic field intensities, which are the only components that could contribute significantly to electromagnetic spin-down. The surface dipolar magnetic field increases by a factor of only three during the first Myr, reaching a maximum a value of $1.5\times 10^{12}$~G, and the quadrupole component remains similarly small, with a maximum of around $6\times 10^{12}$~G. These values are 2 to 3 orders of magnitude smaller than the internal magnetic field strength in the crust. 
\begin{figure}
    \centering
    \includegraphics[width=0.8\columnwidth]{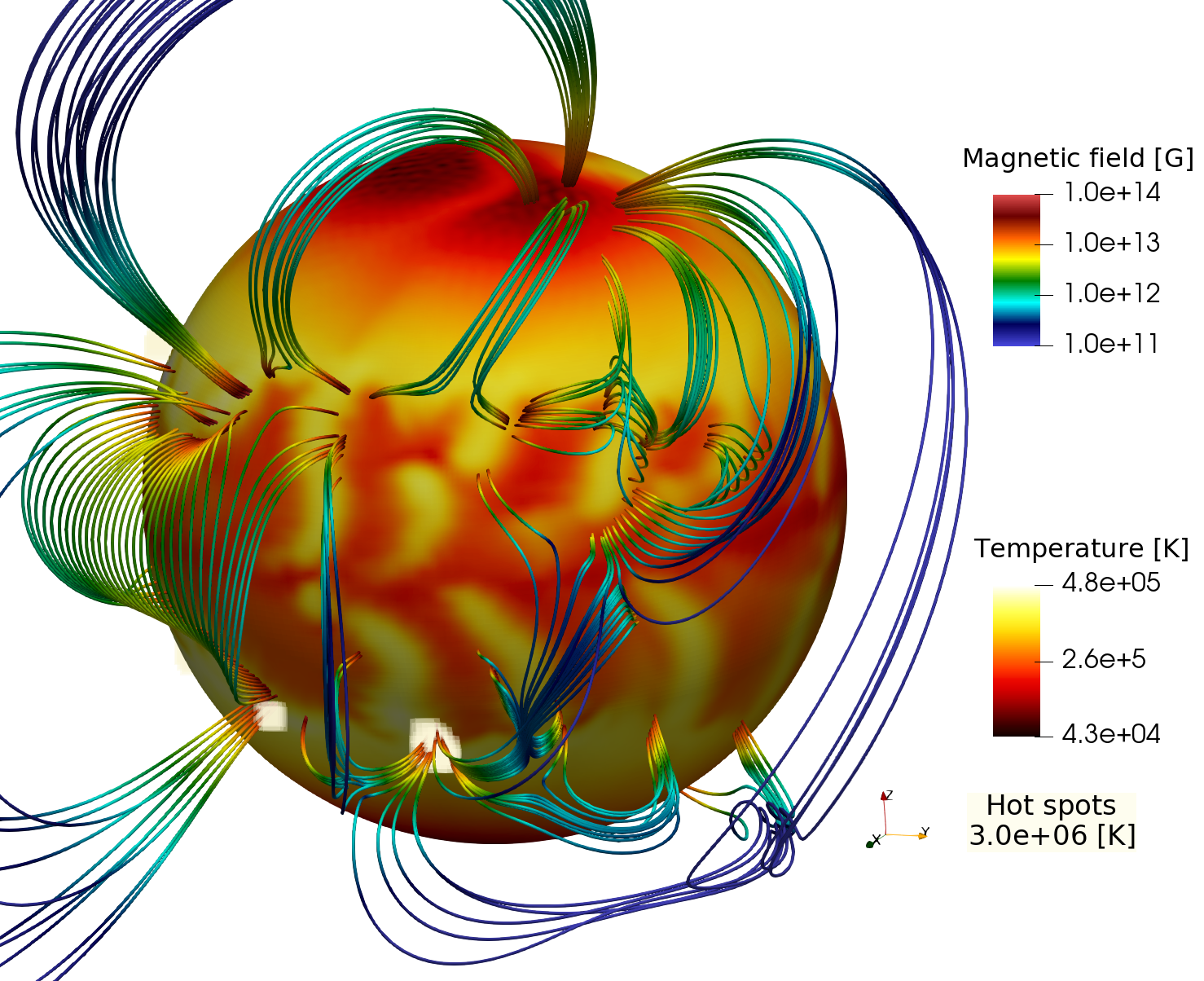}
    \caption{Surface temperature distribution and external magnetic field structure at age 200~kyr.}
    \label{fig:magnetconf_extern}
\end{figure}
Figure~\ref{fig:magnetconf_extern} shows a complex surface magnetic field topology featuring individual arches elongated in the north-south direction. The local field strength at the footpoints of these arches reaches $10^{14}$~G, 100 times stronger than the dipolar magnetic field. Small-scale magnetic fields remain dominant at all times from the beginning of the evolution until 1~Myr (see Methods Section Figure~\ref{fig:spec_evol}). Our numerical simulation therefore successfully reproduces two crucial properties of low-field magnetars: (1) weak dipolar magnetic field, and (2) presence of very strong (50-100 times stronger) small-scale magnetic fields, similar to those found in SGR 0418+5729 \cite{Tiengo2013Natur} and Swift J1882.3-1606 \cite{RodriguezCastillo2016MNRAS}.

\section*{Surface temperatures and hot spots}

The X-ray observations of low-field magnetars are consistent with thermal emission from isolated hot spots with sizes $\leq 1$~km \cite{GuillotPerna2015MNRAS} and black body temperatures reaching  $T_\mathrm{bb}=0.12$~--~$0.6$~keV. The bulk NS emission is not detected with typical upper limits $<10^{31}$~erg/s. SGR 0418+5729 has a pulsed fraction of $62\pm 10$~\% in the [0.3-1.2]~keV range   \cite{GuillotPerna2015MNRAS}. CXOU J164710.2-455216 and Swift J1822.3-1606 have quiescent pulsed fractions $80\pm 3$~\% and $38\pm 3$~\% in the [0.5,10]~keV range respectively \cite{SeoLee2023JKAS}. The upper limit on the bulk thermal emission indicates that low-field magnetars are at least $\approx 200$~kyr old because the bulk X-ray emission drops below $10^{31}$~erg/s after 200~kyr \cite{PotekhinChabrier2018AA} for strongly magnetised NSs (with internal field strengths $\sim 10^{15}$~G).

Strong magnetic fields could create large surface temperature variations, as can be seen in Figure~\ref{fig:magnetconf_extern} (and Methods Section Figure~\ref{fig:T_surf}). We see variations of an order of magnitude between the hottest ($T \approx 4.8\times 10^5$~K) and coldest ($T \approx 4.3\times 10^4$~K) regions. These variations could cause up to 20~\% pulsed fraction but would stay undetectable because of the low bulk X-ray luminosity 
$10^{31}$~erg/s and small effective black body temperature $T_\mathrm{bb} = 0.028$~keV.

We suggest that the observational properties of low-field magnetars can be explained by magnetospheric heating on the small-scale magnetic arches visible in Figure~\ref{fig:magnetconf_extern}. The twisted magnetic field lines penetrate the NS surface through some of these individual footpoints heating surface and forming hot spots. The size of individual footpoints is a fraction of a kilometre, thus emission generated from these footpoints would have properties of emission seen from low-field magnetars, i.e.~very high temperature and small emission area.

Spots are heated by the strongest radial electric currents, which coincide with the strongest radial magnetic fields because $\vec J \propto \vec \nabla \times \vec B = \mu \vec B$ according to the force-free condition in the magnetosphere  \cite{ThompsonLyutikov2002ApJ}. Here we assume that only footpoints with radial magnetic field $|B_r| > 7\times 10^{13}$~G are heated. Under this assumption, it is possible to form up to 10 independent hot spots (see Figure~\ref{fig:magnetconf_extern}) which, if heated to $3\times 10^6$~K, produce luminosity $2\times 10^{32}$~erg/s and emission area with radius $\approx 0.9$~km. The lightcurve is sine-like with a pulsed fraction reaching 92\% for a favourable orientation even without beaming, in agreement with X-ray observations of low-field magnetars (see Methods Section Figures~\ref{fig:light_curve} and \ref{fig:light_curve_fit}). If we increase the critical $|B_r|$ to larger values then we obtain fewer hot spots with smaller areas, and if we decrease the critical $|B_r|$ then the heated area is increased. If the X-ray thermal emission is indeed generated close to the footpoints of these arches, the arches themselves provide natural sites where Compton scattering occurs and absorption features are formed.

\section*{Magnetar bursts}

\begin{figure}
    \centering
    \includegraphics[width=0.9\columnwidth]{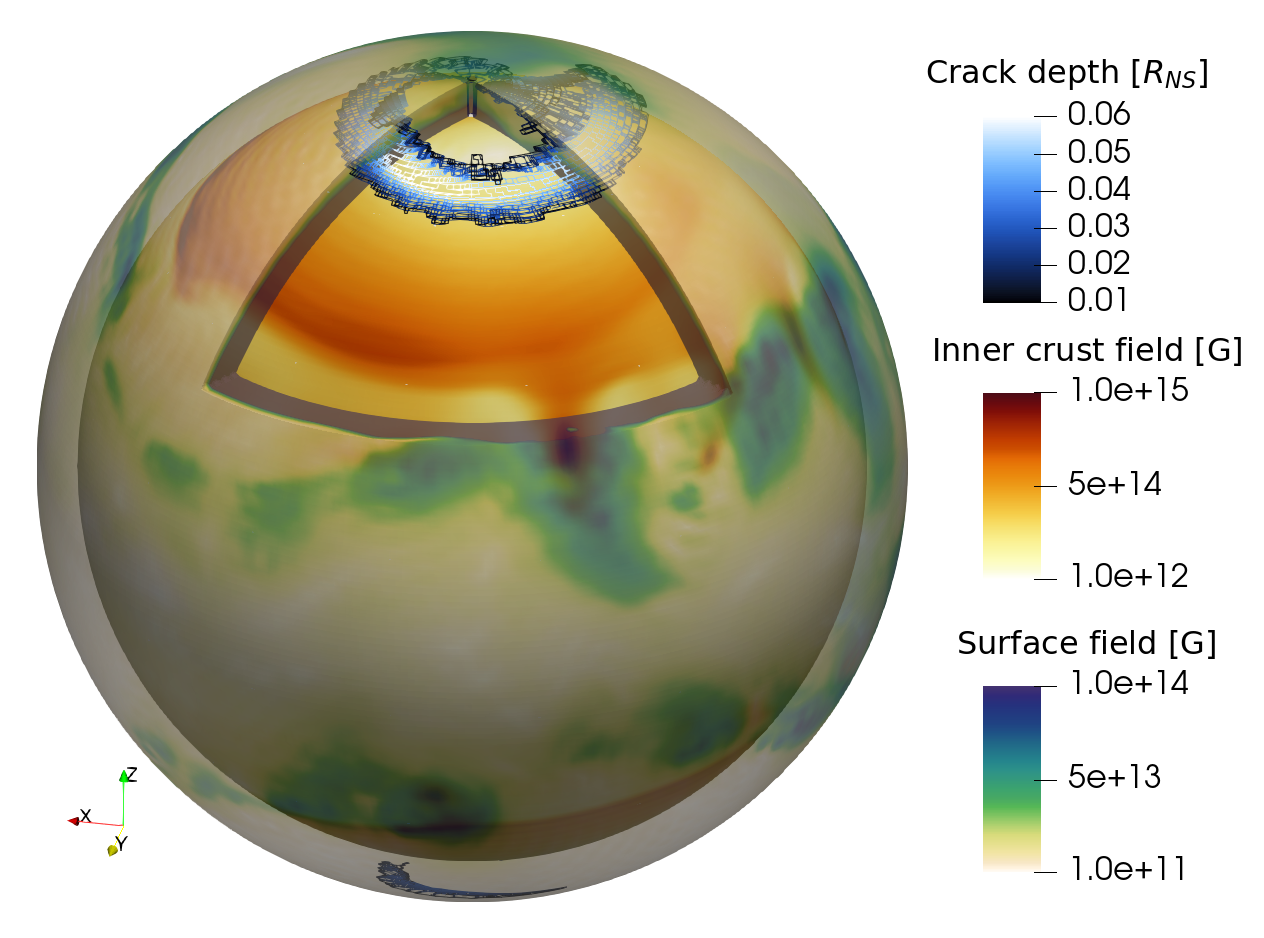}
    \caption{Surface and inner crust magnetic field developed by 200~kyr. Crust yielding regions are shown in white and blue colours.
    }\label{fig:cracks}
\end{figure}

In order to assess whether this magnetic field configuration can power the X-ray activity characteristic of magnetars, we examine the magnetic stresses inside the crust. Bursts and outbursts of magnetars are indeed thought to be caused by crust failure or plastic deformation due to the magnetic stresses \cite{Thompson1995MNRAS,Lander2015MNRAS,Lander2019MNRAS}. We apply the Lander \& Gourgouliatos model \cite{Lander2019MNRAS} and compare the crustal magnetic stresses with the von Mises criterion for crust-yielding (see Methods Section~\ref{sm:failure}). In order to obtain a conservative estimate, the crust is assumed to have completely relaxed only after 2~kyr. Figure~\ref{fig:cracks} shows the average depth of crust failure regions developed at the age of 200~kyr. All the failing regions are located close to the original north and south magnetic poles, coinciding with the regions of strongest magnetic field generated by the proto-NS dynamo. The crust failure regions are much larger in the northern than in the southern hemisphere, due to the properties of the initial magnetic field. This is very different from earlier simulations with simple dipolar initial conditions \cite{Wood2015PhRvL}, in which the crust failure occurred around the original magnetic equator.

In order to check further if the magnetar behaviour could continue at timescales comparable to 100~kyr, we made an additional analysis. We assumed that all the stresses were relaxed in the crust after 100~kyr (which is then used as the reference field $\vec{B}_0$ in Methods Section equation~\ref{eq:crit_stress}), and we compute the magnetic stresses after 200~kyr. Even in this case, the stresses in some crust locations are above the yielding value.

The electromagnetic energy that can potentially be released in such a crustal failure is \cite{Thompson1995MNRAS}
\begin{equation}
E_\mathrm{out} =
4\times 10^{40} \; \mathrm{erg} \left( \frac{l}{1\; \mathrm{km}} \right)^2 \left( \frac{|B|}{10^{15}\; \mathrm{G}} \right)^{2} \approx 2\times 10^{39} \; \mathrm{erg}   
\,,
\end{equation}
where $l\sim 1$~km is the typical size of the failing region and $|B| \sim 2\times 10^{14}$~G (see Figure~\ref{fig:cracks}). 
This value is actually well above the typical burst energy $\sim 10^{37}$~erg of two low-field magnetars: SGR J0418+5729 \cite{vanderHorst2010ApJ} and CXOU J164710.2-455216 \cite{Muno2007MNRAS}. Our modelling provides an upper limit on the extent of crust failure because it maps all the regions which could fail by a certain age. The real size of individual crust failures could be significantly smaller, thus explaining the energy difference between our model and individual bursts observed from magnetars.

\section*{Spin periods}

\begin{figure}
    \centering
    \includegraphics[width=0.8\textwidth]{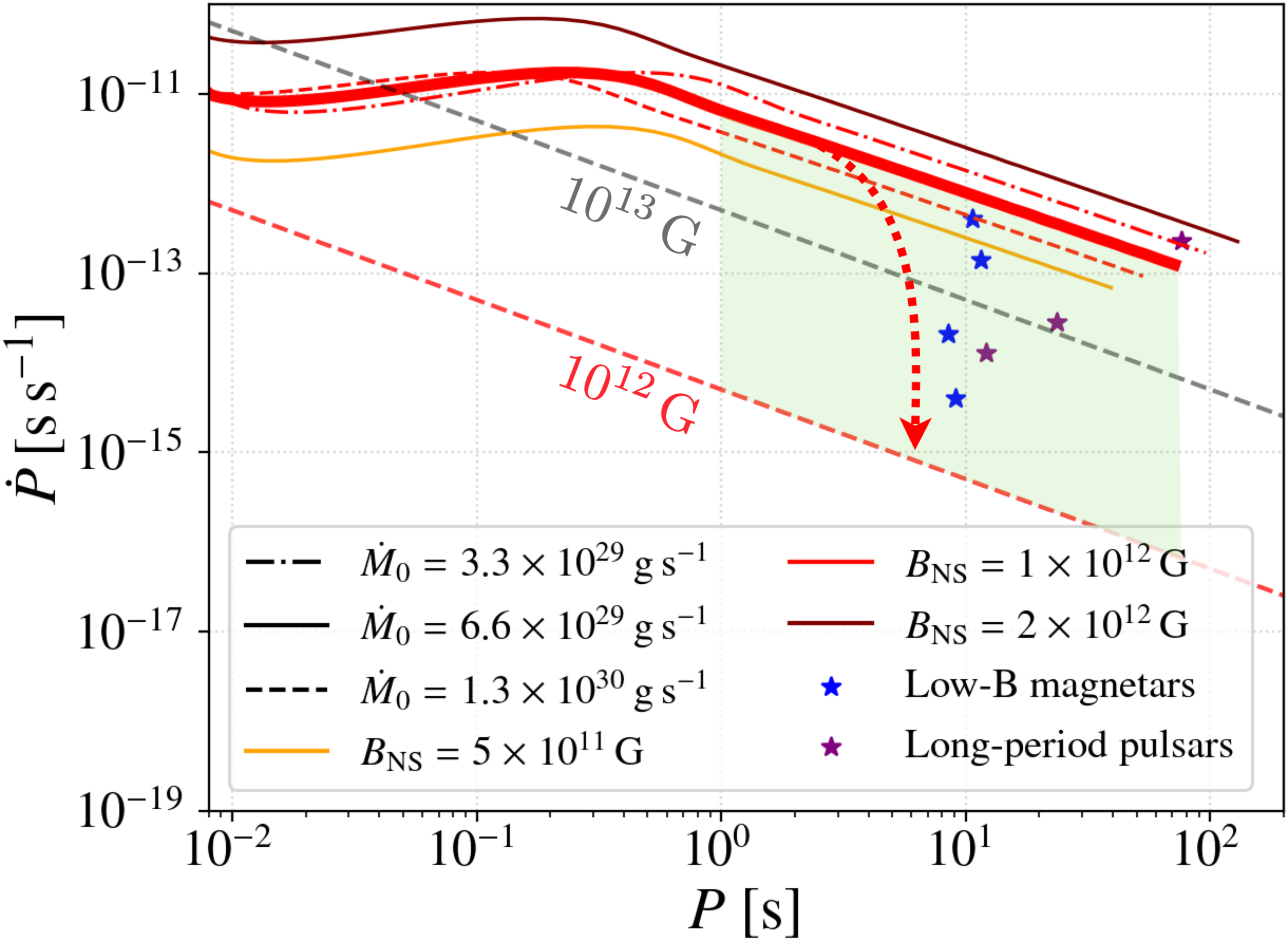}
    \caption{Time evolution of the spin period $P$ and its time derivative $\dot P$ until 10~Myr with small variations of the initial mass accretion rate $\dot M$ and dipolar magnetic field $B_\mathrm{NS}$. The red and grey dashed lines represent the constant dipolar magnetic field lines calculated from the magnetic dipole spin-down formula for $10^{12}$~G and $10^{13}$~G, respectively. The green area covers the zone in which the magnetar could end up if the accretion disk is (partially) depleted (for the fiducial parameters $B_{\rm NS}=10^{12}$~G, $M_{\rm d,0}=0.01$~M$_{\odot}$). The red dotted arrow indicates how the $P\dot{P}$ evolution would behave if the disk is completely depleted.}
    \label{fig:ppdot}
\end{figure}

By electromagnetic braking alone, neutron stars with dipolar fields of a few times $10^{12}$~G cannot reach the spin periods of 8-11~s typical for low-field magnetars on a timescale of 1~Myr. However, it is essential to also take accretion into account, since the Tayler--Spruit dynamo can only develop if the proto-NS accretes fallback material, and this accretion will continue even after the NS is formed. Using the formalism by Ronchi et al. \cite{Ronchi2022ApJ} to model torques from the fallback disk, we naturally obtained periods of 8-11~s after 170~kyr for NSs with dipolar magnetic field similar to our simulations (see Figure~\ref{fig:ppdot}). More details about these calculations are summarised in Methods Section~\ref{s:accretion}. 

Most of the spin-down occurs during the propeller stage when the NS decelerates due to the interaction of its magnetosphere with the fallback disk (see Methods Section Figure~\ref{fig:period}). After 200~kyr, this propeller phase has spun the NS down to a rotation period of $P=8.5$~s and a period derivative  $\dot{P}=\SI{8.5e-13}{s.s^{-1}}$.  According to the standard magnetic dipole spin-down formula, the inferred surface magnetic dipole should then be $B_\text{dip}\approx\SI{3.8e13}{G}$,  which overestimates the true surface magnetic dipole in our simulation by a factor of $\sim 40$. This inferred value of $B_\text{dip}$ is comparable to Swift J1822.3-1606 ($1.4\times 10^{13}$~G) and below the upper limit measured for CXOU J164710.2-455216 ($<6.6\times 10^{13}$~G) as well as 3XMM J185246.6+003317 ($<4\times 10^{13}$~G). 

The apparent magnetic field estimated using instantaneous period and period derivative might be smaller if the disk is partially depleted and provides less torque. Depending on the exact amount of material left in the disk, the period derivative $\dot P$ could range from $\approx 10^{-15}$ (electromagnetic spin-down only) to $\approx 10^{-12}$ (non-depleted disk; green area in Figure~\ref{fig:ppdot}). All low-field magnetars with a measured period derivative fall within this area.

\section*{Impact and future work}

Previous magneto-thermal 
simulations have considered idealised, large-scale magnetic fields \cite{Igoshev2021NatAs,Igoshev2023MNRAS}. Some of these simulations can be made more similar to low-field magnetars by assuming a magnetar-strength dipolar magnetic field which is then dissipated by an increased crust resistivity \cite{Rea2012ApJ} 
\footnote{See e.g. model C0-0-tor with 50\% of toroidal magnetic field and initial dipolar magnetic fields of $10^{14}$~G in \cite{Igoshev2021NatAs}.}. 
Moreover, these simulations stay highly axially symmetric because of the symmetries of the initial conditions. 
Although some previous studies have considered more complicated field structures, as expected from proto-NS evolution \cite{DeGrandis2022ApJ,Dehman2023MNRAS}, our study is the first to directly implement the field from a self-consistent dynamo simulation. The crucial properties of our new magnetic field configuration are that this field is predominantly toroidal and is initially localised near the polar regions of the crust. As a consequence, we find that crustal fractures are most likely to occur in these regions.

One important advance of this work is understanding that the evolution towards large-scale fields is very limited, with the dipole only growing by a factor of two, contrary to earlier suggestions \cite{Sarin2023ApJ}. In our simulation, the magnetic energy contained in the deep crustal fields, including the toroidal component, is significantly larger than the field which can be estimated from surface dipolar values. We find only a moderate increase of the dipolar component, on a timescale of $\sim 10^5$~years.

In comparison to the model suggested by Rea et al. \cite{Rea2012ApJ} with an initial dipolar field as strong as $1.5\times 10^{14}$~G, the dipolar magnetic field stays very low, $\sim 10^{12}$~G, and does not decay significantly in our simulations.

Our results also suggest an important connection between low-field magnetars and recently discovered long-period radio pulsars, such as PSR J0901-4046 \cite{Caleb2022NatAs}.
If the neutron star continues to operate in the propeller phase, it will ultimately reach periods comparable to $75$~s by 10~Myr (Figure~\ref{fig:ppdot}). 
The external magnetic field configuration remains complex, with large open field-line curvature near the NS surface facilitating radio pulsar operation. Thus, pulsar radio emission could occur if the disk is depleted.

Mahlmann et al. \cite{Mahlmann2023ApJ} performed numerical simulations for X-ray outbursts with energies up to $10^{43}$~erg produced by a twisted magnetar magnetosphere. In our simulations, we see the development of individual magnetic arcs and the evolution of their footpoints. Thus, our results can be used as the initial magnetic field for future relativistic magnetosphere simulations.

Our work opens new perspectives for testing extreme dynamos operating in proto-neutron stars. We suggest that different dynamos leave their unique imprint on magnetic field configurations, thus allowing to identify different magnetic amplification processes using the magneto-thermal properties of young isolated neutron stars. While we suggest that the formation of low-field magnetars is linked to the Tayler--Spruit dynamo, the formation of classical magnetars as well as the internal structure of their magnetic fields remains an open question.

\bibliography{pnslib}

\section*{Acknowledgements}
API and RH are supported by STFC grant no.\ ST/W000873/1,
and TSW is supported by STFC grant no.\ ST/W001020/1.
This work was performed using the DiRAC Data Intensive service at Leicester, operated by the University of Leicester IT Services, which forms part of the STFC DiRAC HPC Facility (www.dirac.ac.uk). The equipment was funded by BEIS capital funding via STFC capital grants ST/K000373/1 and ST/R002363/1 and STFC DiRAC Operations grant ST/R001014/1. DiRAC is part of the National e-Infrastructure. This work was also supported by the ``Programme National de Physique Stellaire'' (PNPS) and the ``Programme National des Hautes \'Energies'' (PNHE) of CNRS/INSU co-funded by CEA and CNES. The authors would like to thank the Isaac Newton Institute for Mathematical Sciences, Cambridge, for support and hospitality during the programme DYT2 where work on this paper was undertaken. JG, RR and PB acknowledge support from European Research Council (ERC starting grant no. 715368 – MagBURST) and from the Très Grand Centre de calcul du CEA (TGCC) and IDRIS for providing computational time on Irene and Jean-Zay (allocations A0110410317, A0130410317, A0150410317). 

\section*{Data and materials availability}
For the proto-NS simulation, we used the MagIC code (commit 2266201a5), which is open source at https://github.com/magic-sph/magic. The magnetar spin-down was calculated with the GRB code (commit 84788793), also publicly available at https://github.com/rraynaud/GRBs. The results of magneto-thermal simulations can be shared under reasonable request. 


\section*{Methods section}

\section{Simulation of the proto-neutron star dynamo}
\label{s:protons}

We simulate a proto-NS with a mass of 1.4~$M_{\odot}$ and a radius $R_\mathrm{NS} = \SI{12}{km}$. Its interior is modelled as a stably stratified fluid enclosed between two spherical shells. To control the differential rotation, we impose constant rotation frequencies on both shells (spherical Taylor-Couette configuration), with the outer shell rotating faster than the inner shell to be consistent with the fallback formation scenario. We solve the Boussinesq MHD equations by using the pseudo-spectral code \texttt{MagIC}. In this code, the different lengths $r$, the time $t$, the temperature $T$, and the magnetic field $B$ are scaled as follows:
\begin{equation}
    r \rightarrow rd,\qquad t\rightarrow (d^2/\nu) t,\qquad T\rightarrow (T_o - T_i) T,\qquad B \rightarrow \sqrt{4\pi\rho\eta\Omega_o} B\,,
\end{equation}
with the gap between the two spheres $d=r_o-r_i=\SI{9}{km}$, the kinematic viscosity $\nu=\SI{3.5e9}{cm^2.s^{-1}}$, the temperatures of the outer $T_o$ and inner $T_i$ spheres, the constant density $\rho=\SI{4.1e14}{g.cm^{-3}}$, the resistivity $\eta=\SI{3.5e9}{cm^2.s^{-1}}$, and the rotation rate of the outer sphere $\Omega_o=\SI{628}{rad.s^{-1}}$. So, the dimensionless equations solved by \texttt{MagIC} read
\begin{align} \label{eq:MHD}
    \vec{\nabla} \cdot \vec{v} &=0\,,\\ \label{eq:MHD2}
        \vec\nabla\cdot\vec{B} &=0\,, \\ \label{eq:MHD3}
        \frac{D\vec{v}}{Dt} + \frac{2}{E}\, \vec{e}_z\times \vec{v} &= -\vec{\nabla} p' +\frac{Ra}{Pr}\, T \vec{e}_r 
         + \frac{1}{E\,Pm}\, (\vec\nabla\times \vec{B})\times \vec{B} + \Delta\vec{v}\,, 
    \\ \label{eq:MHD4}
    \frac{DT}{Dt} &=\frac{1}{Pr}\,\Delta T\,,\\ \label{eq:MHD5}
    \frac{\partial\vec{B}}{\partial t} &=\vec\nabla\times(\vec{u}\times\vec{B})+\frac{1}{Pm}\,\Delta \vec{B}\,, 
\end{align}
where $\vec v$ and $\vec B$ are the velocity and magnetic fields, and $p'$ is the non-hydrostatic pressure. $D/Dt\equiv \partial/\partial t + \vec v \cdot \vec \nabla$ is the Lagrangian derivative. $E$, $Ra$, $Pr$, and $Pm$ are dimensionless numbers, which depend on the fluid properties. The Ekman number $E$ is defined as the ratio of the rotation period to the viscous timescale
\begin{equation}
    E = \frac{\nu}{d^2\Omega_o} = \SI{e-5}{}\,.
\end{equation}
The thermal and magnetic Prandtl numbers are defined by
\begin{equation}
    Pr = \frac{\nu}{\kappa} = 0.1
    \qquad \mbox{and} \qquad
    Pm = \frac{\nu}{\eta} = 1\,,
\end{equation}
where $\kappa=\SI{3.5e10}{cm^2.s^{-1}}$ is the thermal diffusivity. Finally, the Rayleigh number $Ra$ measures the ratio between the timescales of thermal transport by diffusion to the thermal transport by convection,
\begin{equation}
    Ra = \left(\frac{N}{\Omega_o}\right)^2\frac{Pr}{E^2}\,,
\end{equation}
where
\begin{equation}\label{eq:brunt-vaisala}
    N\equiv \sqrt{-\frac{g_0}{\rho}\left(\left.\frac{\partial\rho}{\partial S}\right|_{P,Y_e}\frac{dS}{dr}+\left.\frac{\partial \rho}{\partial Y_e}\right|_{P,S}\frac{dY_e}{dr}\right)}
    =\SI{68.2}{s^{-1}}
    \,
\end{equation}
is the Brunt-V{\"a}is{\"a}l{\"a} frequency. The gravitational acceleration is assumed purely radial $\vec{g}=g_0r/r_0\vec{e}_r$. $Y_e$, and $S$ are the electron fraction, and the entropy, respectively.

\par The resolution used is $(n_r,n_{\theta},n_{\phi})=(257,256,512)$. For more information on the numerical methods, see the supplementary materials of~\cite{Barrere2023MNRAS}.

\section{Conversion between \texttt{MagIC} and \texttt{PARODY} codes}
\label{a:conversion}

The poloidal-toroidal decompositions and thus the magnetic potentials are defined differently in the \texttt{MagIC} and \texttt{PARODY} codes. Specifically,
\begin{align}
\vec B &= \vec \nabla \times \vec \nabla \times (b_{\rm pol}^\mathrm{M} \vec e_r) +  \vec \nabla \times (b_{\rm tor}^\mathrm{M} \vec e_r)  \,, \\
\vec B &= \vec \nabla \times \vec \nabla \times (b_{\rm pol}^\mathrm{P} r \vec e_r) +  \vec \nabla \times (b_{\rm tor}^\mathrm{P} r \vec e_r)  
\,,
\end{align}
where the superscript $\mathrm{M}$ / $\mathrm{P}$ refers to \texttt{MagIC} / \texttt{PARODY}, respectively.

Moreover, the codes use different normalisation factors $C_{lm}$ for the spherical harmonics~$Y_l^m (\theta,\phi)$. The spherical harmonics are normalised as the following in the \texttt{PARODY} code
\begin{equation}
C^\mathrm{P}_{lm} = \sqrt{(2-\delta_{m,0})(2l+1) \frac{(l-m)!}{(l+m)!}} \,,
\end{equation}
while the 
normalisation in the \texttt{MagIC} code reads
\begin{equation}
C^\mathrm{M}_{lm} = \frac{1}{1+\delta_{m,0}}\sqrt{\frac{(2l+1)}{4\pi} \frac{(l-m)!}{(l+m)!}} 
\,,
\end{equation}
where $\delta_{m,0}$ is the Kronecker delta, so $\delta_{m,0}= 1$ if $m=0$ and it is 0 otherwise. 

Thus, for the radial magnetic field, we have
\begin{equation}
B_r = \frac{l(l+1)}{r_\mathrm{m}^2}\, b_{\rm pol}^{lm, \rm M} (r_\mathrm{m})\, C_{lm}^\mathrm{M}\, Y_l^m (\theta, \phi) =
\frac{l(l+1)}{r_\mathrm{p}}\, b_{\rm pol}^{lm, \rm P} (r_\mathrm{p})\, C_{lm}^\mathrm{P}\, Y_l^m (\theta, \phi)\,.
\end{equation}
Doing this comparison for each $(l,m)$ separately we thus obtain
\begin{equation}
b_{\rm pol}^{lm, \rm P} (r_\mathrm{pol}) = b_{\rm pol}^{lm, \rm M} (r_\mathrm{m})\, \frac{r_\mathrm{p}}{r_\mathrm{m}^2} \frac{ C_{lm}^\mathrm{M}}{C_{lm}^\mathrm{P}}\,.
\end{equation}
Expanding and simplifying this expression we obtain two different equations for axisymmetric and non-axisymmetric poloidal potentials 
\begin{align}
b_{\rm pol}^{l0, \rm P} (r_\mathrm{p}) &=  
\frac{b_{\rm pol}^{l0, \rm M} (r_\mathrm{m})}{\sqrt{4 \pi}}
\frac{r_\mathrm{p}}{r_\mathrm{m}^2} 
\quad{\rm for}\quad m=0\,,\\ 
b_{\rm pol}^{lm, \rm P} (r_\mathrm{p}) &=  
\frac{b_{\rm pol}^{lm, \rm M} (r_\mathrm{m})}{\sqrt{2 \pi}}
\quad{\rm for}\quad m\neq0\,.
\end{align}

Similarly, we can proceed with the $\theta$-component of the magnetic field computed using only the toroidal potential
\begin{equation}
B_\theta = \frac{C_{lm}^\mathrm{M}}{r_\mathrm{m}\sin\theta}\, b_{\rm tor}^{lm, \rm M}(r_\mathrm{m})\, \frac{\partial Y_l^m (\theta,\phi) }{\partial \phi}  =  \frac{C_{lm}^\mathrm{P}}{\sin\theta}\, b_{\rm tor}^{lm, \rm P}(r_\mathrm{p})\, \frac{\partial  Y_l^m (\theta,\phi)}{\partial \phi}\,.
\end{equation}
Thus, the normalisation is
\begin{equation}
b_{\rm tor}^{lm, \rm P} (r_\mathrm{p}) =  b_{\rm tor}^{lm, \rm M} (r_\mathrm{m})\, \frac{C_{lm}^\mathrm{M}}{C_{lm}^\mathrm{P}} \frac{1}{r_\mathrm{m}}\,,
\end{equation}
which simplifies to
\begin{align}
b_{\rm tor}^{l0, \rm P} (r_\mathrm{p}) &=
\frac{b_{\rm tor}^{l0, \rm M} (r_\mathrm{m})}{\sqrt{4\pi} }
\frac{1}{r_\mathrm{m}}
\quad{\rm for}\quad m=0\,,\\
b_{\rm tor}^{lm, \rm P} (r_\mathrm{p}) &=
\frac{b_{\rm tor}^{lm, \rm M} (r_\mathrm{m})}{\sqrt{2\pi} }
\frac{1}{r_\mathrm{m}}
\quad{\rm for}\quad m\neq0\,.
\end{align}

\begin{figure}
    \centering
    \includegraphics[width=0.7\textwidth]{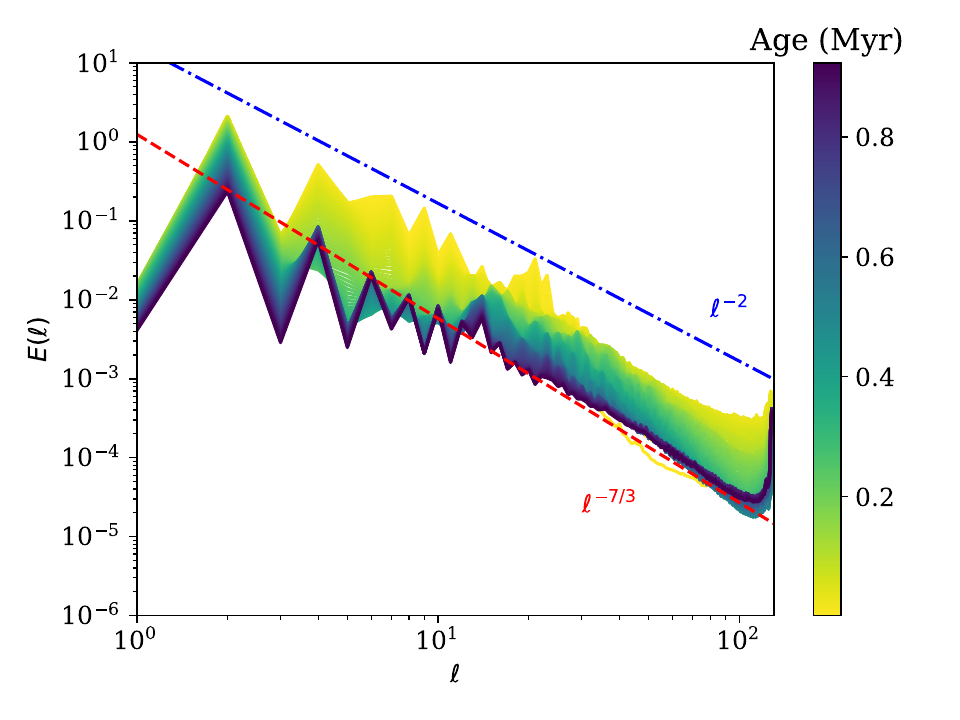}
    \caption{Evolution of the magnetic energy spectra over 1~Myr.}
    \label{fig:spec_evol}
\end{figure}

In this work, we preserve the angular structure obtained in dynamo simulations at the surface and in the middle of the crust up to $L_\mathrm{max} = 30$, which corresponds to surface structures of $\approx 1$~km. Analysis of the dynamo simulations reveal that larger-scale structures do indeed dominate the magnetic field. Smaller scale structures are generated during the first kyr via the Hall cascade, see Figure~\ref{fig:spec_evol}.

\section{Crust-confined magnetic field configurations}
\label{sm:configuration}

In addition to the technical details in the previous section, the proto-NS dynamo setup and the magneto-thermal crust evolution setup differ in their geometry, having aspect ratios $\chi_{\rm pNS}=0.25$ and $\chi_{\rm NS} = 0.9$, respectively. Thus, in order to create a magnetic field configuration which is similar to proto-NS results but is also crust-confined, we should extract only the top 10\% of the proto-NS simulation.

Our approach for importing the results of the dynamo simulations is to require all components of the magnetic field to be exactly the same at certain points within the crust. We consider the poloidal and toroidal potentials for each individual spherical harmonic, and require both these potentials to exactly coincide with our numerical fits at the following points: $r_1 = 0.93$ and $r_2 = 0.96$. We require our fit for the poloidal potential to coincide at the surface. We also require our poloidal and toroidal potentials to satisfy the potential boundary condition at the surface and the `no-currents' boundary condition at the core-crust interface.

Similarly to recent work \cite{Igoshev2023MNRAS} we represent the radial part of the poloidal and toroidal potentials as a polynomial expansion
\begin{equation}
b_{lm} (r) = \frac{a_0 + a_1 r + a_2 r^2 + a_3 r^2 + a_4 r^4 }{r} \,.
\end{equation}
Overall, all conditions for the radial part of the poloidal potential can be written as
\begin{equation}
\begin{array}{ccc}
  b_p (1)   & = & \beta_p (1.0)\,, \\
  b_p (0.96)   & = & \beta_p (0.96)\,, \\
  b_p (0.93)   & = & \beta_p (0.93)\,, \\
  b_p (r_c) & = & 0\,, \\
  \frac{\partial b_p}{\partial r} (1) + \frac{(l+1)}{r} b_p (1) &= & 0\,.
\end{array} \label{eq:cond_pol}   
\end{equation}
Here $\beta_p(r)$ are coefficients of the spectral expansion for poloidal magnetic field extracted from the proto-NS \texttt{MagIC} simulations.
These conditions are individually satisfied for each $l$ and $m$, and translate into the following system of linear equations
\begin{equation}
\begin{array}{lcl}
a_0 + a_1 + a_2 + a_3 + a_4  & = & \beta_p(1.0)\,,  \\
(a_0 + a_1 r_1 + a_2 r_1^2 + a_3 r_1^3 + a_4 r_1^4 ) / r_1 & = & \beta_p(r_1) \,,\\
(a_0 + a_1 r_2 + a_2 r_2^2 + a_3 r_2^3 + a_4 r_2^4 ) / r_2 & = & \beta_p(r_2) \,,\\
a_0 + a_1 r_c + a_2 r_c^2 + a_3 r_c^3 + a_4 r_c^4  & = & 0\,, \\
a_0 l + (l+1) a_1 + (l+2) a_2 + (l+3) a_3 + (l+4) a_4  & = & 0\,.\\
\end{array}
\end{equation}

For the toroidal potential we use the following conditions
\begin{equation}
\begin{array}{lcl}
  b_t (1)   & = & 0\,, \\
  b_t (0.96)   & = & \beta_t (0.96)\,, \\
  b_t (0.93)   & = & \beta_t (0.93)\,, \\
  \partial \left[r b_t (r_c) \right] / \partial r & = & 0\,. \\
\end{array} \label{eq:cond_tor}   
\end{equation}
Similarly, $\beta_t(r)$ here are the coefficients of the spectral expansion for the toroidal magnetic field extracted from the proto-NS simulations.
These conditions then translate into the linear system
\begin{equation}
\begin{array}{lcc}
a_0 + a_1 + a_2 + a_3  & = & 0\,,  \\
(a_0 + a_1 r_1 + a_2 r_1^2 + a_3 r_1^3 ) / r_1 & = & \beta_t(r_1)\,, \\
(a_0 + a_1 r_2 + a_2 r_2^2 + a_3 r_2^3 ) / r_2 & = & \beta_t(r_2)\,, \\
a_1 + 2 a_2 r_c + 3 a_3 r_c^2 = 0\,.\\
\end{array}
\end{equation}

\section{Simulation of neutron star magneto-thermal evolution}
\label{sm:magnetothermalsimulations}

The pseudo-spectral code \texttt{PARODY} \cite{dormy1997} was modified to solve 
the following
system of dimensionless partial differential equations for magnetic field $\vec B$ and temperature $T$:
\begin{equation}
\frac{\partial \vec B}{\partial t} = \mathrm{Ha}\, \vec \nabla \times \left[ \frac{1}{\mu^3}\vec B \times (\vec\nabla \times \vec B)\right] - \vec \nabla \times \left[\frac{1}{\mu^2} \vec \nabla \times \vec B \right] + \mathrm{Se}\, \nabla \left[\frac{1}{\mu}\right]\times \vec \nabla T^2\,,
\end{equation}
\begin{equation}
\frac{\mu^2}{\mathrm{Ro}}\, \frac{\partial T^2}{\partial t} = \vec \nabla \cdot \left[\mu^2 \hat \chi \cdot \vec \nabla T^2 \right] + \frac{\mathrm{Pe}}{\mathrm{Se}}\, \frac{|\vec \nabla \times \vec B|^2}{\mu^2} + \mathrm{Pe}\, \mu \left[ \vec \nabla \times \vec B \right] \cdot \vec \nabla \left[ \frac{T^2}{\mu^2}\right]\,,
\end{equation}
where the first equation is the magnetic induction equation and the second is the thermal diffusion equation. The terms on the right-hand side of the first equation correspond to the Hall effect, Ohmic decay and the Biermann battery effect. The terms on the right-hand side of the second equation correspond to anisotropic thermal diffusion, Ohmic heating and entropy carried by electrons. The derivation of the above equations is summarised in \cite{DeGrandis2020ApJ}. The same code was also used to compute the evolution of off-centred dipole configurations \cite{Igoshev2023MNRAS}.

The electron chemical potential varies within the crust as
\begin{equation}
\mu(r) = \mu_0 \left[ 1 + \frac{(1 - r/R_\mathrm{NS})}{ 0.0463} \right]^{4/3}\,.
\end{equation}
The tensor $\hat \chi$ describing the anisotropy of the heat transport is written as:
\begin{equation}
\hat \chi = \frac{\delta_{ij} + \mathrm{Ha}\, B_i B_j / \mu^2 - \mathrm{Ha}\, \epsilon_{ijk} B_k / \mu}{1 + \mathrm{Ha}^2\, |\vec B|^2/\mu^2}\,,
\end{equation}
where $\delta_{ij}$ is the Kronecker symbol and $\epsilon_{ijk}$ is the Levi-Civita symbol.

The dimensionless Hall ($\mathrm{Ha}$), Seebeck ($\mathrm{Se}$), P\'{e}clet ($\mathrm{Pe}$) and Roberts ($\mathrm{Ro}$) parameters depend on the chosen scales for the magnetic field and temperature, which we take to be $B_0 = 10^{14}$~G and $T_0 = 1.0\times 10^{8}$~K.
The Hall number is defined by
\begin{equation}
\mathrm{Ha} = c \tau_0 \frac{e B_0}{\mu_0} \approx 49.1    
\,,
\end{equation}
where $e$ is the electron charge, $c$ is the speed of light, $\tau_0 = 9.9\times 10^{19}$~s is the electron scattering relaxation time \cite{UrpinYakovlev1980SvA} and $\mu_0 = 2.9\times 10^{-5}$~erg is the electron chemical potential at the top of the crust.
The Seebeck number is defined by
\begin{equation}
\mathrm{Se} = 2\pi^3 k_\mathrm{B}^2 T_0^2 n_0 e \frac{c\tau_0}{\mu_0 B_0} \approx 0.052   
\,,
\end{equation}
where $k_\mathrm{B}$ is the Boltzmann constant and $n_0 = 2.603\times 10^{34}$~cm$^{-3}$ is the electron number density at the top of the crust.
Finally, the P\'{e}clet and Roberts numbers are
\begin{equation}
\mathrm{Pe} = \frac{3}{4\pi} \frac{B_0}{e n_0 c \tau_0} \approx 6.44\times 10^{-5}
\,,
\end{equation}
and
\begin{equation}
\mathrm{Ro} = \frac{3}{4\pi^3} \frac{\mu_0^2}{k_\mathrm{B}T_0} \frac{1}{c^2 \tau_0^2} \frac{1}{e^2 n_0} \approx 3580 
\,.
\end{equation}

In order to ensure the solenoidality of the magnetic field $\vec B$, we write the magnetic field as a sum of poloidal and toroidal parts
\begin{equation}
\vec B = \vec \nabla \times  \vec \nabla \times (b_\text{pol} \vec r)  + \vec \nabla \times (b_\text{tor} \vec r)\,.
\end{equation}
The scalar potentials $b_\text{pol}$ and $b_\text{tor}$ are expanded in spherical harmonics.
We model the core as a perfect conductor,
which implies the following inner boundary conditions at $r=0.9$
\begin{equation}
b_\text{pol} = 0
\qquad \mbox{and} \qquad
\frac{d(rb_\text{tor})}{dr} = 0\,.
\end{equation}
We model the region outside the NS as a vacuum,
which implies the following outer boundary conditions at $r=1$
\begin{equation}
\frac{db_\text{pol}^{lm}}{dr} + \frac{l+1}{r} b_\text{pol}^{lm} = 0
\qquad \mbox{and} \qquad
b_\text{tor} = 0\,,
\end{equation}
where $b_\text{pol}^{lm}$ is the coefficient of degree $l$ and order $m$ in the spherical harmonic expansion of the poloidal potential $b_\text{pol}$.

The temperature is fixed to its initial value at the core--crust boundary (see more details about modelling cooling at the end of the section). The outer boundary condition for the temperature is
\begin{equation}
-\mu^2 \vec r \cdot \hat \chi \cdot \vec \nabla (T^2) = \frac{1}{5} \frac{R_\text{NS}}{c\tau_0}\; \mathrm{Se} \, \mathrm{Pe}\, (T_s/T_0)^4
\,,
\end{equation}
where the (dimensional) surface temperature $T_s$ is related to the crustal temperature $T_b$ as:
\begin{equation}
\left[ \frac{T_s}{10^6\; \mathrm{K}} \right]^2 = \left[\frac{T_b}{10^8\; \mathrm{K}} \right]
\,,
\end{equation}
using simplified relation \cite{Gudmundsson1983ApJ}.

\begin{figure}
    \centering
    \includegraphics[width=0.7\textwidth]{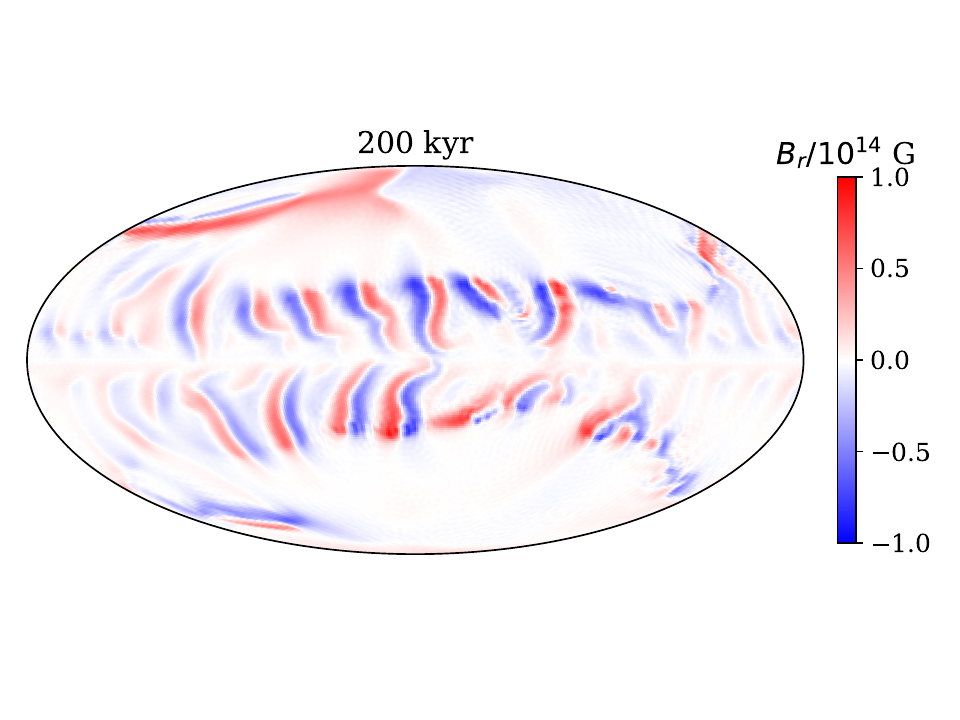}
    \caption{The surface radial magnetic field after 200~kyr of evolution.}
    \label{fig:Br_surf}
\end{figure}

\begin{figure}
    \centering
    \includegraphics[width=0.7\textwidth]{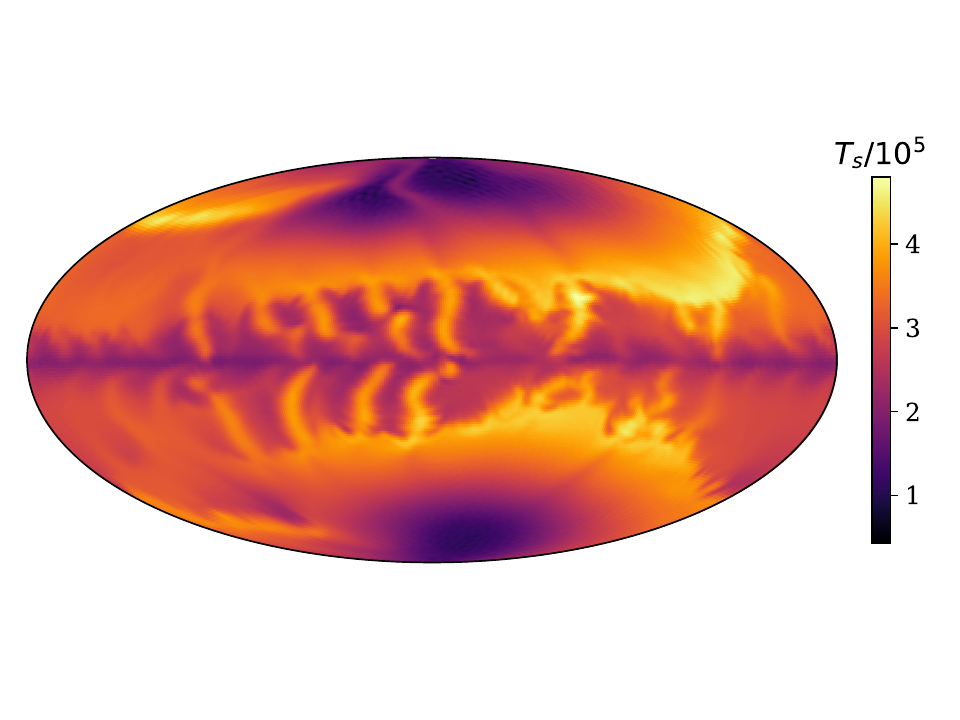}
    \caption{The surface temperature after 200~kyr of evolution assuming a NS core temperature $10^6$~K. No magnetospheric hot spots are shown.}
    \label{fig:T_surf}
\end{figure}

The numerical resolution is $n_r=96$ grid points in the radial direction and spherical harmonic degrees up to $l_\mathrm{max} = 128$.
We show the surface radial magnetic field as well as the surface temperature at age 200~kyr in Figures~\ref{fig:Br_surf} and \ref{fig:T_surf}.
We show the evolution of magnetic energy spectra in Figure~\ref{fig:spec_evol}.

To take into account the NS cooling, we restart calculations at 200~kyr changing the core temperature to $10^6$~K. We run calculations for 1~kyr to allow the simulation to relax, i.e.\ crust temperatures stop evolving on short timescales, creating a stable surface thermal pattern.

\section{Properties of thermal emission}

\begin{figure}
    \centering
    \includegraphics[width=0.7\textwidth]{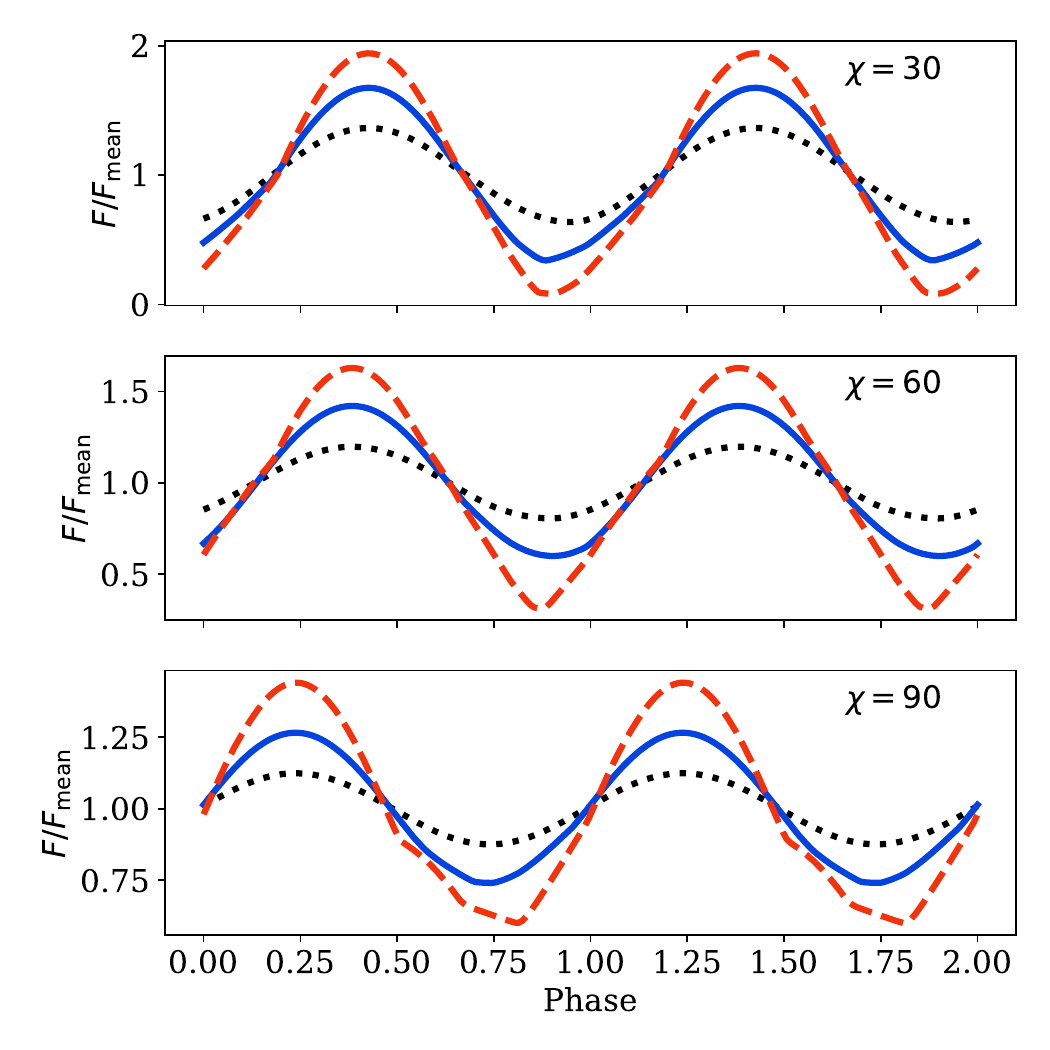}
    \caption{Soft X-ray lightcurves for the surface thermal map combined with the magnetospheric hot spots. Each panel corresponds to a different obliquity angle $\chi$. The three curves correspond to different inclination angles: black dotted lines are for $i=30^\circ$, blue solid lines are for $i=60^\circ$, and red dashed lines are for $i=90^\circ$.}
    \label{fig:light_curve}
\end{figure}

We use the open source code \texttt{Magpies}\footnote{https://github.com/ignotur/magpies} to model X-ray thermal lightcurves. We show these results in Figure~\ref{fig:light_curve}. The maximum pulsed fraction reaches 93 \% for the most favourable orientation of the rotational axis with respect to the original dipole axis.

\begin{figure}
    \centering
    \begin{minipage}{0.49\linewidth}
    \includegraphics[width=0.99\textwidth]{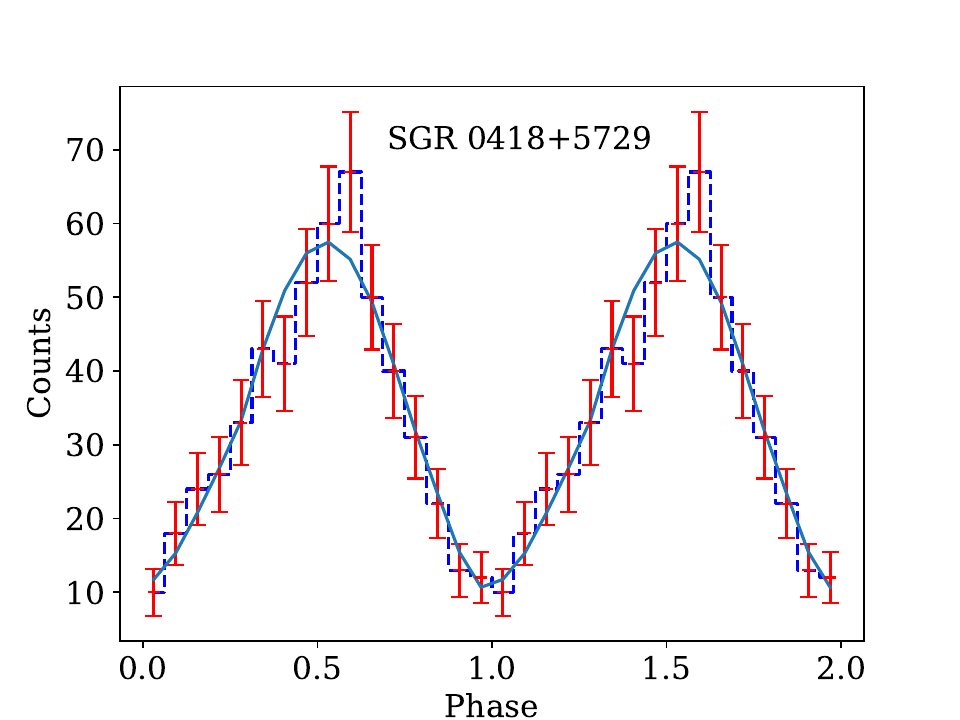}    
    \end{minipage}
    \begin{minipage}{0.49\linewidth}
    \includegraphics[width=0.99\textwidth]{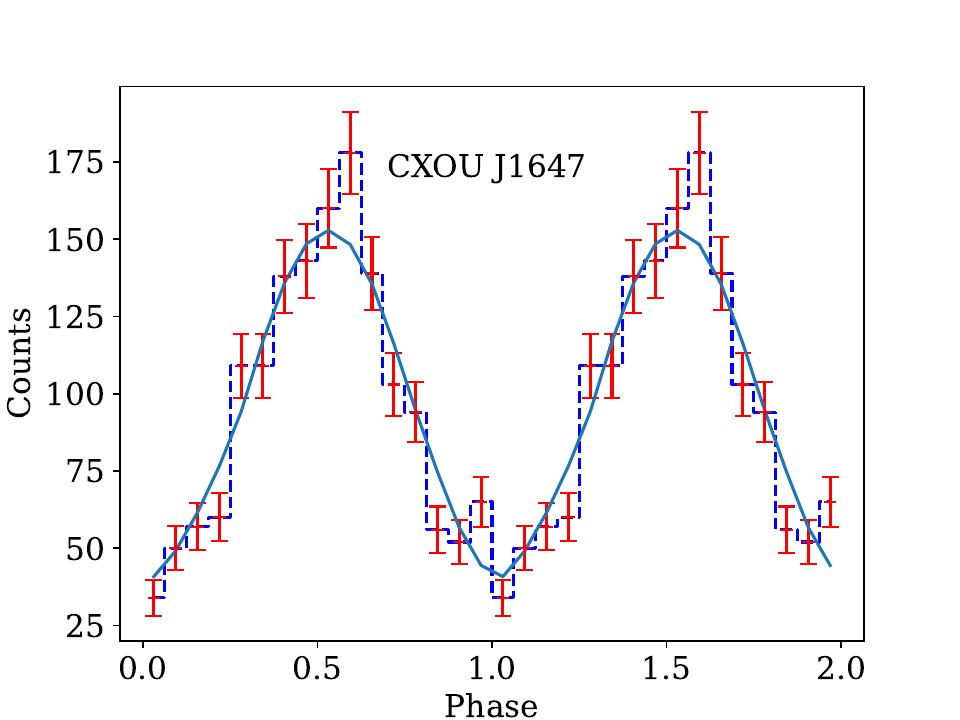}    
    \end{minipage}
    \begin{minipage}{0.49\linewidth}
    \includegraphics[width=0.99\textwidth]{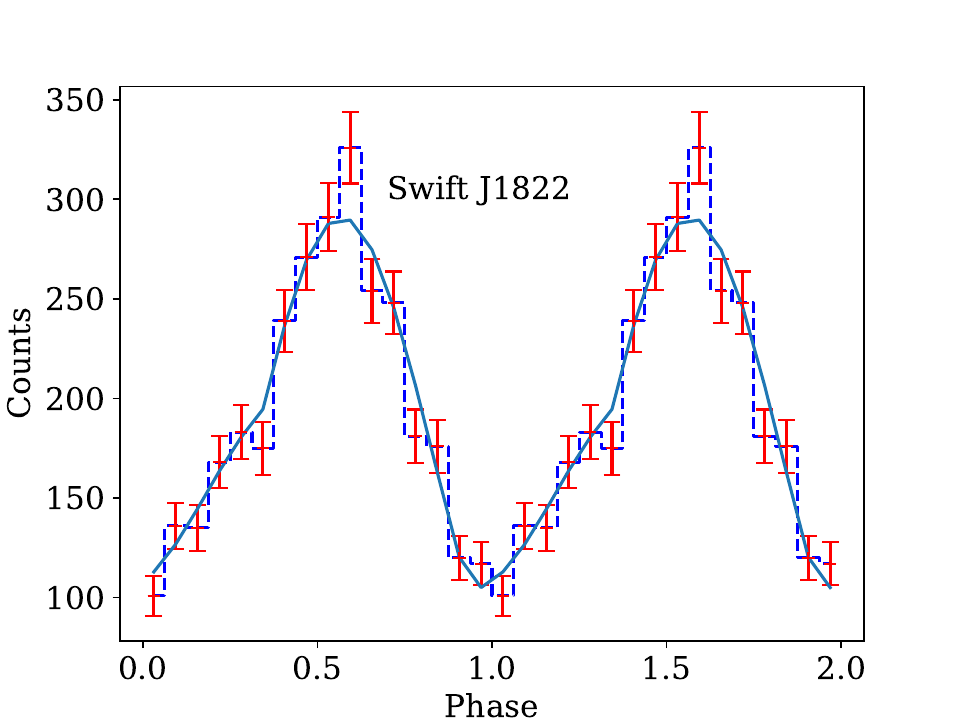}    
    \end{minipage}
    \begin{minipage}{0.49\linewidth}
    \includegraphics[width=0.99\textwidth]{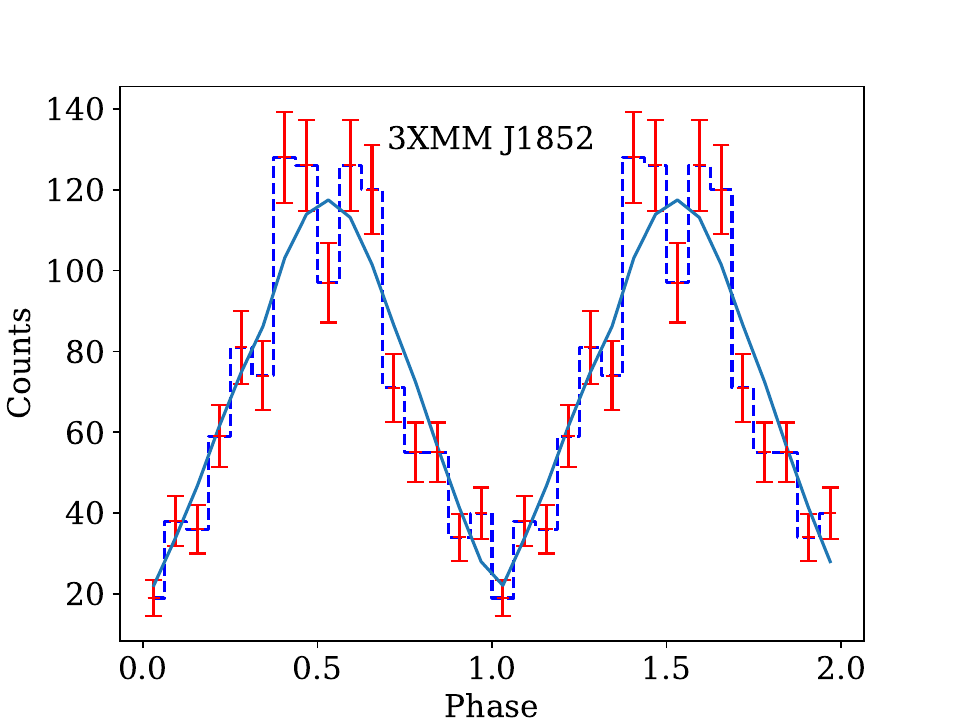}    
    \end{minipage}
    \caption{Observed soft X-ray lightcurves in the range of 0.3-2~keV for low-field magnetars in the quiescent state and the best fits (solid blue line). It is assumed that emission is produced by hot spots formed at the places with radial magnetic field exceeding $7\times 10^{13}$~G.}
    \label{fig:light_curve_fit}
\end{figure}

Similarly to \cite{Igoshev2021NatAs} we try to fit the soft X-ray lightcurve in the range 0.3-2~keV. We show the results in Figure~\ref{fig:light_curve_fit}. We summarise the obliquity angle as well as inclination angles in Table~\ref{tab:my_label}. While SGR 0418+5729 and Swift J1822.3-1606 are fitted relatively well, the two remaining magnetars have more features in the lightcurves.

\begin{table}
    \centering
    \begin{tabular}{lcccc}
    \hline
    low-field magnetar         & $\chi$ & $i$   & $\Delta \Phi$ & C-stat \\
                           & (rad)  & (rad) & (rad)         \\
    \hline
    SGR 0418+5729          & 0.6984 & 1.264 & 5.616 & 6.8  \\
    CXOU J164710.2-455216  & 0.7518 & 1.085 & 5.555 & 29.9 \\
    Swift J1822.3-1606     & 0.0519 & 0.636 & 5.625 & 16.5 \\
    3XMM J185246.6+003317  & 1.093  & 1.637 & 5.330 & 34.7 \\
    \hline
    \end{tabular}
    \caption{Possible rotational orientation for low-field magnetars assuming that their X-ray lightcurves are produced by hot spots.}
    \label{tab:my_label}
\end{table}

\section{Crust failure}
\label{sm:failure}

We use here a model developed by \cite{Lander2019MNRAS} based on earlier work by \cite{Lander2015MNRAS}. Essentially, we use the von Mises criterion for crust-yielding following Eq.\ (14) of \cite{Lander2019MNRAS}:
\begin{equation}
\tau_\mathrm{el} \leq \frac{1}{4\pi}\sqrt{\frac{1}{3}\vec B_0^4 + \frac{1}{3}\vec B^4 + \frac{1}{3}\vec B_0^2 \vec B^2 - (\vec B\cdot \vec B_0)^2}\,.
\label{eq:crit_stress} 
\end{equation}
Here $\vec B_0$ is the relaxed (initial) state of the magnetic field, which we assume to coincide with our first simulation snapshot at $2$~kyr. $\tau_\mathrm{el}$ is the scalar yield stress. The field $\vec B$ is computed at 200~kyr. We compute the critical strain following the procedure by \cite{Lander2019MNRAS} with a correction (private communication
by Sam Lander)
\begin{equation}
\tilde\rho = 99.6 \left(1 - \frac{R_\mathrm{cc}}{R_\mathrm{nd}}\right)^2  (1 - \mathcal R)^2 + 0.004\,,
\end{equation}
where $\mathcal R$ is computed as:
\begin{equation}
\mathcal{R} = \frac{r - R_\mathrm{cc}}{R_\mathrm{nd} - R_\mathrm{cc}}\,,
\end{equation}
where $R_\mathrm{cc} = 0.9$ is the location of the crust-core interface and $R_\mathrm{nd} = 1$ is the location of the neutron-drip point.  
Thus, our critical strain varies from $\approx 8\times 10^{26}$~g~cm$^{-1}$~s$^{-2}$ close to the neutron-drip boundary to $4.6\times 10^{29}$~g~cm$^{-1}$~s$^{-2}$ at the core-crust boundary. Following our normalisation, the stress caused by Lorentz forces (right-hand side of equation~\ref{eq:crit_stress}) is multiplied by a numerical factor $(10^{14}\; \mathrm{G})^2$. This von Mises criterion is written assuming that failure occurs in the form of shearing motion \cite{Lander2019MNRAS}.

\section{Accretion driven spin-down}
\label{s:accretion}
To explain the NS spin-down to the regime of low-field magnetars, we invoke the propeller mechanism due to the interaction between the NS magnetic field and the remaining fallback disk. The evolution of the NS-fallback depends on the three different radii: (i) the light cylinder radius, (ii) the magnetospheric radius, and (iii) the corotation radius, which are defined by the respective expressions
\begin{equation}
    r_{\rm lc} = \frac{c}{\Omega_{\rm NS}}\,,
\end{equation}
\begin{equation}
    r_{\rm mag} = \mu^{4/7}(GM_{\rm NS})^{-1/7}\dot{M}^{-2/7}\,,
\end{equation}
\begin{equation}
    r_{\rm cor} = \left(\frac{GM_{\rm NS}}{\Omega_{\rm NS}^2}\right)^{1/3}\,.
\end{equation}
Here $c$ is the speed of light, and $G$ is the gravitational constant. $M_{\rm NS}$ and $\Omega_{\rm NS}$ are the NS mass and rotation rate; $\mu=B_{\rm NS}R_{\rm NS}^3$ is its magnetic dipole moment. $\dot{M}$ is the accretion rate\footnote{Strictly speaking, $\dot{M}$ is the material loss rate from the accretion disk. In the propeller regime this quantity remains positive even though the material is not accreted onto the neutron star.}.

If the disk penetrates the magnetosphere ($r_{\rm lc}>r_{\rm mag}$), it can either spin up the NS by accreting matter if $r_{\rm cor}>r_{\rm mag}$, or spin down the NS  in a propeller phase if $r_{\rm cor}<r_{\rm mag}$. In this propeller phase, the magnetic field accelerates the inner disk to super-Keplerian speeds, which produces a centrifugal outflow. Angular momentum is therefore transported from the NS toward the disk, which can efficiently spin down the NS.

\begin{figure}
    \centering
    \includegraphics[width=0.7\textwidth]{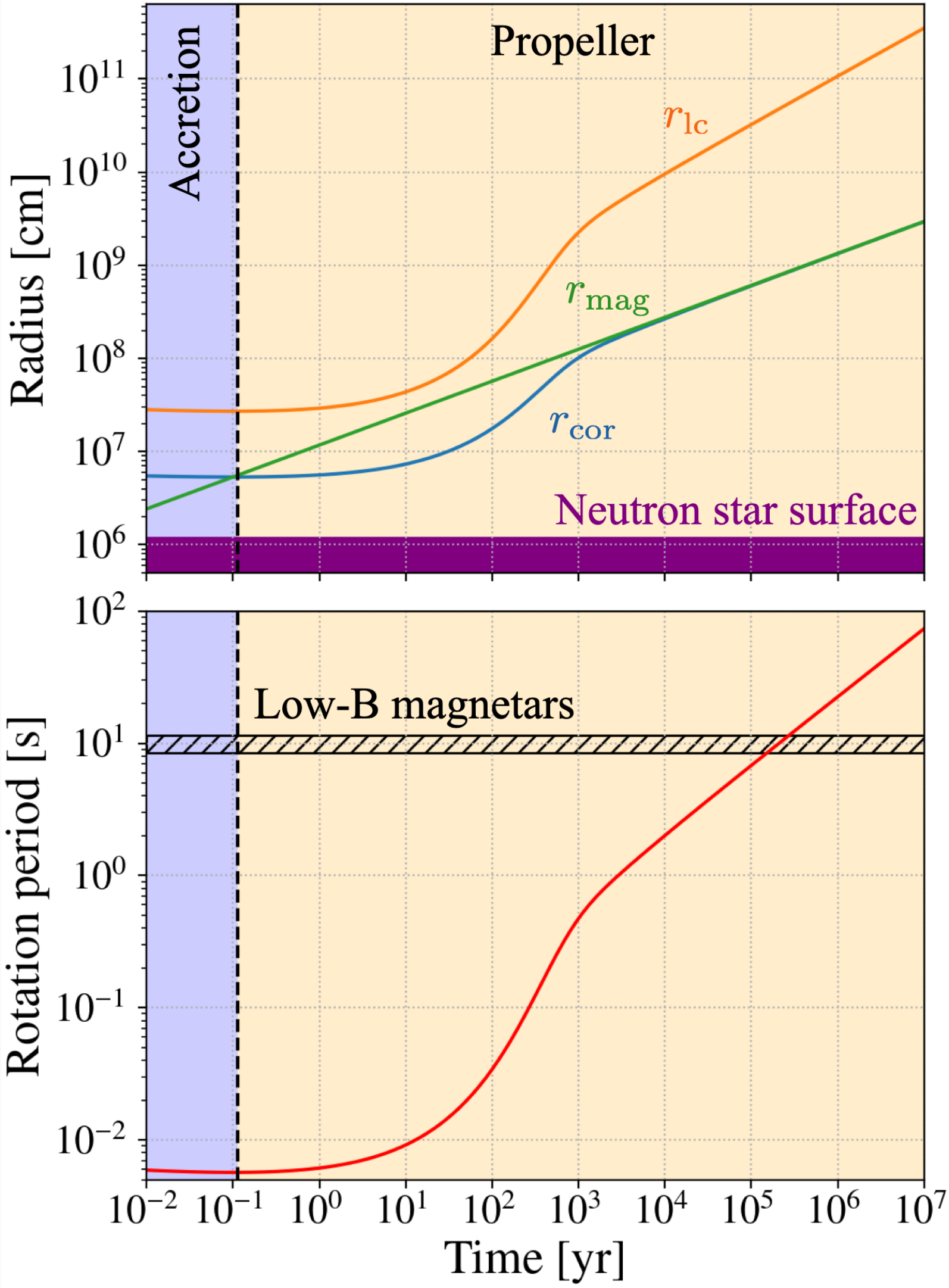}
    \caption{Time evolution of the characteristic radii (top) and the NS rotation period (bottom) for $B_{\rm NS}=10^{12}$~G, $M_{\rm d,0}=0.01$~M$_{\odot}$, and an initial rotation period of 10~ms. The NS is spun up during the accretion regime (blue region) and strongly spun down in the propeller phase (orange region). The hatched region in the bottom figure represents the range of rotation periods observed in low-field magnetars.}
    \label{fig:period}
\end{figure}

The modelling of the NS-fallback evolution we use is strongly inspired by~\cite{Gompertz2014MNRAS} except for the mass accretion rate, which reads~\cite{Ronchi2022ApJ}
\begin{equation}
    \dot{M}(t)=\dot{M}_0\left(1+\frac{t}{t_{\nu}}\right)^{-1.2}\,,
\end{equation}
where $t_{\nu}\sim 30$~s is the viscous timescale and $\dot{M}_0=M_{\rm d,0}/t_{\nu}\sim 6.5\times10^{29}$~g$\,$s$^{-1}$ is the initial accretion rate, and $M_{\rm d,0}=\SI{0.01}{M_{\odot}}$ is the initial fallback disk mass. The torques exerted on the NS by the accretion disk are given by
\begin{equation}
    N_{\rm acc}=
    \begin{cases}
        \left(1-\left(\frac{r_{\rm mag}}{r_{\rm cor}}\right)^{3/2}\right)\sqrt{GM_{\rm NS}R_{\rm NS}\dot{M}^2} & \: {\rm if} \: r_{\rm mag}>R_{\rm NS}\,,\\
        \left(1-\left(\frac{\Omega_{\rm NS}}{\Omega_{\rm K}}\right)^{3/2}\right)\sqrt{GM_{\rm NS}R_{\rm NS}\dot{M}^2} & \: {\rm if} \: r_{\rm mag}<R_{\rm NS}\,,
    \end{cases}
\end{equation}
where $\Omega_{\rm K}=\sqrt{GM_{\rm NS}/R_{\rm NS}^3}$ is the Keplerian angular velocity. The dipole spins the NS down as follows
\begin{equation}
    N_{\rm dip}=-\frac{2}{3}\frac{\mu^2\Omega_{\rm NS}^3}{c^3}\left(\frac{r_{\rm lc}}{r_{\rm mag}}\right)^3\,.
\end{equation}
Therefore the NS angular velocity evolves as
\begin{equation}\label{eq:torque}
    I_{\rm NS}\dot{\Omega}_{\rm NS}=N_{\rm acc} + N_{\rm dip}\,,
\end{equation}
where $I_{\rm NS}=1.45\times10^{45}$~g$\,$cm$^2$ is the NS moment of inertia. Figure~\ref{fig:period} shows the time series of the characteristic radii and NS rotation period that result from the solution of equation~\eqref{eq:torque} for $B_{\rm NS}=10^{12}$~G, $M_{\rm d,0}=0.01$~M$_{\odot}$, and an initial rotation period of 10~ms. We clearly find that the NS is strongly spun down during the propeller phase and reaches the period range of the observed low-field magnetars at $\sim 170$~kyr. This timescale varies up to $\sim 550$~kyr for $B_{\rm NS}=5\times10^{11}$~G. Figure~\ref{fig:period} shows the period and period derivative evolution. 

\end{document}